\documentstyle[emulateapj,psfig]{article}

\makeatletter

\newenvironment{inlinefigure}{%
\def\@captype{figure}%
\noindent\begin{minipage}{0.999\linewidth}\begin{center}}
{\end{center}\end{minipage}\smallskip}
\makeatother

\lefthead{Barger et al.}

\begin{document}
\title{The Submillimeter Properties of the 1~Ms {\it Chandra} 
Deep Field North X-ray Sample}
\author{A.\,J.\ Barger,$\!$\altaffilmark{1,2,3}
L.\,L.\ Cowie,$\!$\altaffilmark{1}
A.\,T.\ Steffen,$\!$\altaffilmark{2}
A.\,E.\ Hornschemeier,$\!$\altaffilmark{4}
W.\,N.\ Brandt,$\!$\altaffilmark{4}
G.\,P.\ Garmire$\!$\altaffilmark{4}
}

\altaffiltext{1}{Institute for Astronomy, University of Hawaii,
2680 Woodlawn Drive, Honolulu, Hawaii 96822}
\altaffiltext{2}{Department of Astronomy, University of Wisconsin-Madison, 
475 North Charter Street, Madison, WI 53706}
\altaffiltext{3}{Hubble Fellow and Chandra Fellow at Large}
%\altaffiltext{4}{Visiting Astronomer, W.\,M.\ Keck Observatory,
%jointly operated by the California Institute of Technology and
%the University of California}
\altaffiltext{4}{Department of Astronomy \& Astrophysics,
525 Davey Laboratory, The Pennsylvania State University,
University Park, PA 16802}

\slugcomment{Submitted to the Astrophysical Journal Letters}

\begin{abstract}

We present submillimeter observations for 136 of the 
370 X-ray sources detected in the 1~Ms exposure 
of the {\it Chandra} Deep Field North. Ten of the 
X-ray sources are significantly detected in the submillimeter.
The average X-ray source in the sample has a significant
850~$\mu$m flux of $1.69\pm 0.27$~mJy. This value shows
little dependence on the $2-8$~keV flux from 
$5\times 10^{-16}$~erg~cm$^{-2}$~s$^{-1}$
to $10^{-14}$~erg~cm$^{-2}$~s$^{-1}$.
The ensemble of X-ray sources contribute about 10\% of the 
extragalactic background light at 850~$\mu$m. The submillimeter 
excess is found to be strongest in the optically faint X-ray sources 
that are also seen at 20~cm, which is consistent with these X-ray 
sources being obscured and at high redshift ($z>1$).

\end{abstract}


\section{Introduction}
\label{secintro}

Most of the energy density of the extragalactic X-ray background
(XRB) resides in the hard X-ray band above 2~keV.
Early $100-300$~ks {\it Chandra} observations, in combination 
with {\it ASCA} observations at the bright flux levels, 
resolved $>60$\% of the $2-10$~keV XRB into discrete sources
(\markcite{mushotzky00}Mushotzky et al.\ 2001;
\markcite{giacconi01}Giacconi et al.\ 2001;
\markcite{garmire01a}Garmire et al.\ 2001a;
\markcite{tozzi01}Tozzi et al.\ 2001). 
Now two unprecedented 1~Ms {\it Chandra} exposures
(\markcite{brandt01}Brandt et al.\ 2001;
\markcite{rosati01}Rosati et al.\ 2001) have essentially completely
resolved the remainder of the XRB at these hard energies.  

Many of the hard X-ray sources in the ultradeep {\it Chandra}
data appear to be highly absorbed systems with large hard to 
soft X-ray flux ratios. In these sources the rest-frame soft X-ray 
through near-infrared radiation is reprocessed by dust and gas 
surrounding the central active galactic nucleus (AGN),
and the reradiated energy appears in the far-infrared (FIR).
At high redshifts ($z>1$) this FIR radiation
is redshifted to the submillimeter
and can be directly measured with the SCUBA camera 
(\markcite{holland99}Holland et al.\ 1999) on the 15~m 
James Clerk Maxwell Telescope.

Early searches for submillimeter counterparts to {\it Chandra} 
X-ray sources yielded mixed results because of the small sample
sizes: \markcite{fabian00}Fabian et al.\ (2000; A2390 and A1835) 
and \markcite{horn00}Hornschemeier et al.\ (2000, 2001; 
Hubble Deep Field North and its vicinity)
found no submillimeter counterparts to their hard X-ray sources, 
while \markcite{bautz00}Bautz et al.\ (2000; A370)
found submillimeter counterparts to both of theirs. 
\markcite{barger01a}Barger et al.\ (2001a) used submillimeter data 
that covered the entire 57~arcmin$^2$ SSA13 {\it Chandra} X-ray 
field of \markcite{mushotzky00}Mushotzky et al.\ (2000) to look for
overlaps between the hard X-ray and submillimeter populations. 
Although only one hard X-ray source was significantly detected 
in their submillimeter data, these authors found that the 
error-weighted sum of the submillimeter fluxes of all 20 hard X-ray 
sources in their sample was significant at $20\pm 6$~mJy and mostly 
arose from the $z>1.5$ and spectroscopically unidentified sources.

In this paper we present submillimeter observations of
the X-ray sources detected in the 1~Ms exposure of the 
{\it Chandra} Deep Field North (CDF-N). These data
cover more than three times the area of the 
\markcite{barger01a}Barger et al.\ (2001a)
submillimeter/X-ray data. Furthermore, the 1~Ms hard X-ray data 
go about a factor of 10 deeper than the 100~ks SSA13 data, providing 
a much larger sample with a wide dynamic range in X-ray flux. 
The 1~Ms X-ray source catalog presented in
\markcite{brandt01}Brandt et al.\ (2001) contains
370 significantly detected point sources.
%360 in the $0.5-8$~keV (full) band,
%325 in the $0.5-2$~keV (soft) band, 265 in the $2-8$~keV
%(hard) band, and 145 in the $4-8$~keV (ultrahard) band.
Source positions are accurate to within $\approx 0.6-1.7''$,
depending on the off-axis angle and the number of detected
counts. For our analysis, we adopt from Table~3 of 
\markcite{brandt01}Brandt et al.\ (2001) the effective photon 
index ($\Gamma$) values, which are 
based on the ratio of the counts in the hard and soft bands,
and the observed frame hard band flux values.
The $R$ magnitudes and redshifts are taken from 
the 1~Ms X-ray follow-up catalog of 
\markcite{barger01b}Barger et al.\ (2001b),
which gives redshift identifications for 66 of the 
136 X-ray sources with submillimeter observations.

\section{Submillimeter Observations}
\label{secsmm}

SCUBA jiggle map observations at 850~$\mu$m
were taken during observing 
runs in 2000 April and 2001 March and May.
The maps were dithered to prevent any regions of the sky from
repeatedly falling on bad bolometers. The chop throw was
fixed at a position angle of 90~deg so that the negative beams
would appear $45''$ on either side east-west of the positive beam.
Regular ``skydips'' (\markcite{manual}Lightfoot et al.\ 1998)
were obtained to measure the zenith atmospheric opacities,
and the 225\ GHz sky
opacity was monitored at all times to check for sky stability.
Pointing checks were performed every
hour during the observations on the blazars 0923+392, 
0954+685, 1144+402, 1216+487, 1219+285, 1308+326, or 1418+546.
The data were calibrated using jiggle maps of the primary 
calibration source Mars or the secondary calibration sources 
CRL618, OH231.8+4.2, or 16293-2422.
Submillimeter fluxes were measured using beam-weighted extraction 
routines that include both the positive and negative portions of 
the beam profile.

The data were reduced in a standard and consistent way using 
the dedicated SCUBA User Reduction Facility
(SURF; \markcite{surf}Jenness \& Lightfoot 1998).
The new data were combined in the reduction process 
with the jiggle maps previously 
obtained by \markcite{bcr00}Barger, Cowie, \& Richards (2000).
Due to the variation in the density of bolometer samples across
the maps, there is a rapid increase in the noise levels at the
very edges, so the low exposure edges were clipped.

The SURF reduction routines arbitrarily normalize all the data
maps in a reduction sequence to the central pixel of the first
map; thus, the noise levels in a combined image are
determined relative to the quality of the central pixel in the
first map. In order to determine the absolute noise levels of
our maps, we first eliminated the $\gtrsim 3\sigma$ real sources 
in each field by subtracting an appropriately normalized version
of the beam profile. We then iteratively adjusted the noise
normalization until the dispersion of the signal-to-noise
ratio measured at random positions became $\sim 1$. The noise 
estimate includes both fainter sources and correlated noise.

\section{Submillimeter Flux Measurements}
\label{smmflux}

The submillimeter fluxes of the X-ray sources were measured
in an automated fashion. We first measured the submillimeter 
fluxes at the positions of the 20~cm ($>3\sigma$) sources from
\markcite{richards00}Richards (2000) that fell on our 
SCUBA maps. Any radio source that we detected above the $3\sigma$ 
level at 850~$\mu$m was included in a ``detection list''.
We next measured all $>5\sigma$ 
submillimeter sources that did not have a radio counterpart;
we identified two and added them to the list. 
Finally, we added all the 
\markcite{chapman01}Chapman et al.\ (2001) submillimeter
measurements (made in photometry mode) of radio sources
that did not lie within our SCUBA jiggle maps.

After compiling the detection list, we compared the coordinates
of the X-ray sources with the coordinates of the sources in the 
detection list. If an X-ray source was found to lie within $3''$ 
of a source in the detection list, we identified the list source 
as the counterpart to the X-ray source.
All of the radio+submillimeter source counterparts had coordinates
within $1.5''$ of the X-ray source positions. Once the 
cross-identifications were made, the sources in the detection 
list were removed from the SCUBA maps since the wide and complex 
beam patterns of the bright SCUBA sources can produce spurious
detections at other positions. 
Submillimeter fluxes and uncertainties were then measured at all 
the unassigned X-ray positions using a recursive loop which 
removed X-ray sources that were detected in the submillimeter 
above the $3\sigma$ level in a descending 
order of significance prior to remeasuring the fluxes. This 
procedure again avoids multiple or spurious detections of a 
single bright submillimeter source. 

Of the 370 {\it Chandra} X-ray sources in the CDF-N sample, 136
have 850~$\mu$m flux measurements with uncertainties less than
5~mJy and 109 have uncertainties less
than 2.5~mJy. Five flux measurements were drawn from the
Chapman et al.\ (2001) sample, including a significant
$15.7\pm 2.4$~mJy detection. While we did not reproduce the
submillimeter detection of one of the Chapman et al.\ sources
which fell within our jiggle maps (VLA~J123624+621743 was
observed by Chapman et al.\ in poor weather conditions and 
found to be a marginal $3\sigma$ detection (see their Table~1); 
the source was also detected at the $3.7\sigma$
level in the scanmap of \markcite{borys01}Borys et al.\ 2001),
the very high significance of the 15.7~mJy source suggests 
that it is real. Complete
exclusion of the Chapman et al.\ data has no significant
effect on our subsequent analysis.

In total, we significantly ($>3\sigma$) detect in the 
submillimeter 10 of the 136 X-ray sources. The average 
submillimeter flux per X-ray source is 1.30~mJy, and the
error-weighted average is $1.69\pm 0.16$~mJy.
The final submillimeter measurements are presented in the CDF-N 
1~Ms X-ray follow-up catalog of 
\markcite{barger01b}Barger et al.\ (2001b) and
are plotted versus hard X-ray flux (squares) in Figure~\ref{fig1}.
We include in the figure the SSA13 hard X-ray sources
(circles) from \markcite{barger01a}Barger et al.\ (2001a)
to better sample the bright, hard X-ray flux end.

%
% Figure 1
%
\begin{inlinefigure}
\psfig{figure=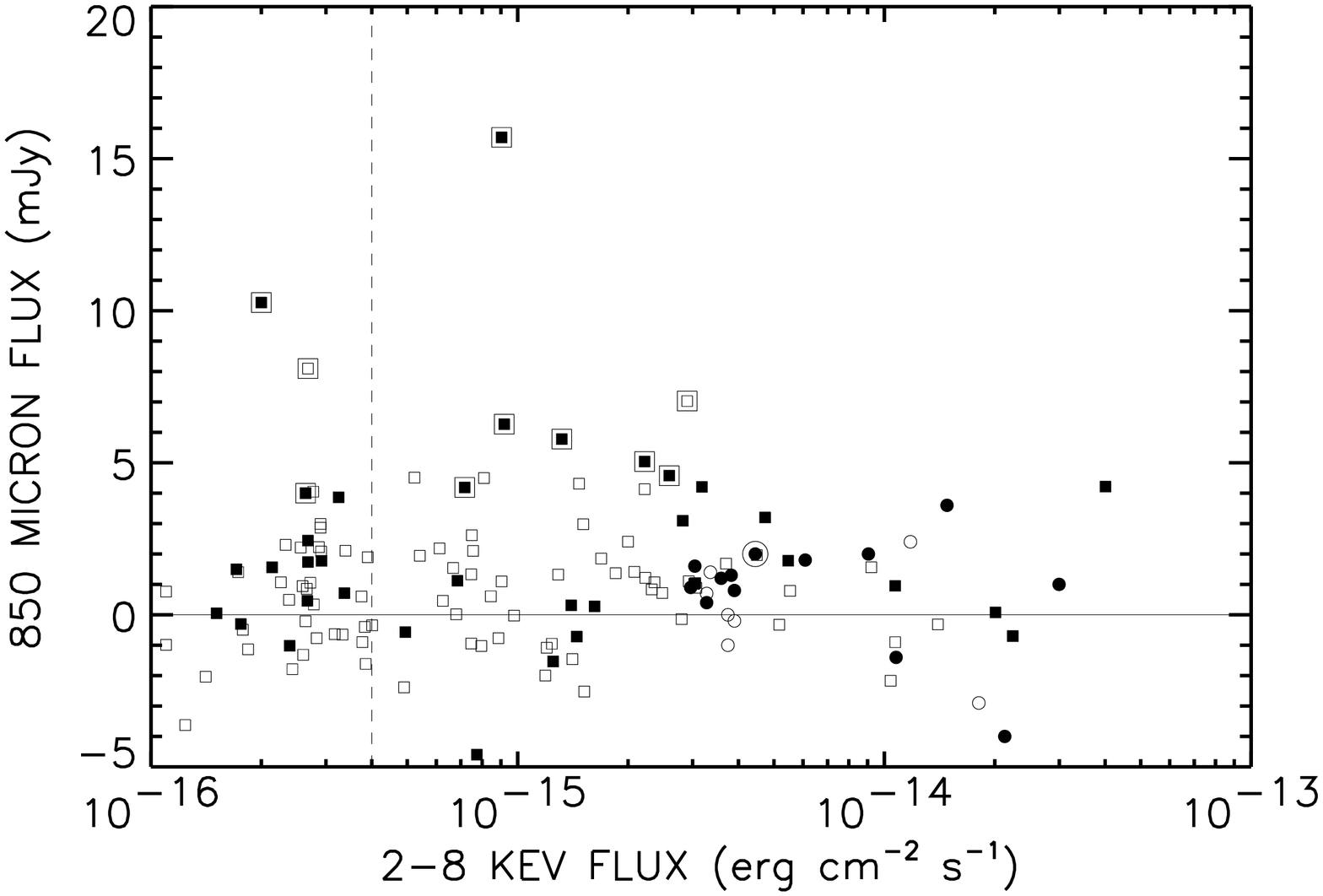,angle=90,width=3.5in}
\vspace{6pt}
\figurenum{1}
\caption{
Submillimeter versus hard X-ray ($2-8$~keV) fluxes for the X-ray
sources in the CDF-N (squares) and SSA13 (circles) fields with
submillimeter measurements.
Sources with $>20\mu$Jy radio fluxes are denoted by filled symbols,
and sources with $>3\sigma$ submillimeter fluxes are denoted by a 
second large symbol. Thirty of the 46 sources to the left of the
dashed vertical line at $4\times 10^{-16}$~erg~cm$^{-2}$~s$^{-1}$
(including the two significant submillimeter sources that are
also radio sources) are plotted at their hard X-ray flux limits.
Two of these have hard
X-ray flux limits less than $10^{-16}$~erg~cm$^{-2}$~s$^{-1}$
and are shown at a nominal value of
$1.1\times 10^{-16}$~erg~cm$^{-2}$~s$^{-1}$.
Only one source to the right of the vertical line has only
a hard X-ray flux limit
($1.25\times 10^{-15}$~erg~cm$^{-2}$~s$^{-1}$).
}
\label{fig1}
\addtolength{\baselineskip}{10pt}
\end{inlinefigure}

In order to test the significance of our results, we repeated
our submillimeter measurement procedure for a large number 
of simulations. In the simulations we offset the X-ray sample a 
random amount in the RA and Dec directions and then measured the 
submillimeter fluxes at these random positions.
The average flux per source at these random positions was found
to be accurately zero, but the dispersion from field to field was
found to be 0.27~mJy rather than the nominal 0.16~mJy, suggesting
the presence of some systematic error. We use this larger uncertainty
subsequently. The 95 percent confidence level is 0.45~mJy, and
the 98 percent confidence level is 0.6~mJy, leaving little doubt
that there is a significant excess 850~$\mu$m flux associated with
the X-ray sample. Similarly, the random samples have an average
of 1.2 $3\sigma$ sources per field and 95 and 98 percent confidence
levels of 4 and 5 sources, respectively, so most of the
significantly detected submillimeter sources are real.

\section{Contribution to the Submillimeter EBL}

The resolution of the FIR/submillimeter extragalactic background 
light (EBL) by the DIRBE and FIRAS experiments on {\it COBE} revealed
that the FIR/submillimeter EBL has approximately the same 
integrated energy density as the optical EBL. Thus, the 
absorption and reradiation of light by dust in the history of 
galaxy formation and evolution is extremely important; however, 
in order to reconstruct the star formation and accretion 
histories of the Universe, we need to be able to differentiate 
between the FIR/submillimeter contributions made by starbursts and 
those made by AGN.

About $20-30$ percent of the 850~$\mu$m EBL has been 
resolved into discrete sources by blank field SCUBA surveys
(e.g., \markcite{bcs99}Barger, Cowie, \& Sanders 1999).
\markcite{blain99}Blain et al.\ (1999) estimated from a
SCUBA survey of massive cluster lenses which
amplify distant submillimeter sources that more than
80 percent of the 850~$\mu$m EBL is now resolved.
The ultraluminous SCUBA sources are believed to be the 
high redshift analogs of the local ultraluminous infrared 
galaxies (ULIGs; \markcite{sanders96}Sanders \& Mirabel 1996), 
but there is an ongoing debate even as to whether the local ULIGs 
are dominantly powered by star formation or by AGN activity. 
Modelling suggests that $\gtrsim 10-20$ percent of the 
FIR/submillimeter EBL may be contributed by galaxies 
containing bright AGN 
(\markcite{almaini99}Almaini, Lawrence, \& Boyle 1999;
\markcite{gunn01}Gunn \& Shanks 2001).
It is difficult to spectroscopically look for AGN signatures
in the SCUBA sources ---
though in a few cases such signatures have been detected
(e.g., \markcite{ivison98}Ivison et al.\ 1998; 
\markcite{barger99}Barger et al.\ 1999) --- due to the extremely
faint nature of the vast majority of the optical counterparts. 
However, since most of the flux density from the {\it Chandra}
X-ray sample is expected to arise from AGN activity,
we may use the present data to place an upper limit on the AGN 
contribution to the 850~$\mu$m EBL. (A submillimeter
source detected in X-rays is not necessarily solely powered
by accretion onto the AGN, as there may also be dust obscured
starburst activity in the galaxy which contributes to the 
submillimeter flux.)

\markcite{barger01a}Barger et al.\ (2001a) multiplied the
hard XRB measurement from \markcite{vecchi99}Vecchi et al.\ (1999) 
by the ratio of the total 850~$\mu$m flux to the total hard X-ray 
flux of the SSA13 hard X-ray sample to estimate the hard X-ray 
contribution to the 850~$\mu$m EBL. 
They determined the contribution to be about 10 percent. 
However, their approach assumed that a fixed submillimeter 
to X-ray flux ratio applies to all of the X-ray sources.
We can quantitatively see that this is not a good assumption
from Table~\ref{tab1}, where we list the average X-ray flux and 
the average submillimeter flux and uncertainty for the CDF-N+SSA13 
X-ray sources in each of four hard X-ray flux bins.
Since the average submillimeter flux is consistent with being 
constant as a function of X-ray flux, the ratio of the 
submillimeter to X-ray flux rises with decreasing X-ray flux. 

A better approach may be to translate the measured average 
850~$\mu$m flux per X-ray source to an X-ray source 
contribution to the 850~$\mu$m EBL.
The SCUBA jiggle maps cover an area of 194~arcmin$^2$. 
Excluding the \markcite{chapman01}Chapman et al.\ (2001) 
sources, which lie outside this area, there are 131 
X-ray selected sources with an average flux per source 
of $1.27\pm 0.27$~mJy,
where the uncertainty is again based on the random samples
described in \S~\ref{smmflux}.
This translates to an 850~$\mu$m EBL of 
$3.1\pm 0.66\times 10^3$~mJy~deg$^{-2}$, which
is $10\pm 2$ percent of the total 850~$\mu$m EBL, 
if we adopt the $3.1\times 10^4$~mJy~deg$^{-2}$ value of 
\markcite{puget96}Puget et al.\ (1996), or 
$7\pm 2$ percent of the total,
if we adopt the $4.4\times 10^4$~mJy~deg$^{-2}$ 
value of \markcite{fixsen98}Fixsen et al.\ (1998).
Since our results indicate that the integrated 850~$\mu$m
EBL is proportional to the total number of X-ray sources,
the overall contribution to the EBL may increase with additional 
weaker X-ray sources. However, the rapid convergence of the X-ray 
number counts below $10^{-15}$~erg~cm$^{-2}$~s$^{-1}$ 
(\markcite{garmire01b}Garmire et al.\ 2001b)
suggests that this correction would not be large.

\section{Source properties}
  
If the submillimeter flux represents the reprocessing
of the radiated energy in the more obscured AGN, then
we may expect to see a dependence of the submillimeter 
excess on X-ray hardness. However, an opposing 
tendency is introduced by redshift: 850~$\mu$m source 
detections are primarily expected to be at $z>1$ because 
of the strong negative {\it K}-corrections at this wavelength
that nearly compensate for cosmological dimming, while 
high redshift X-ray sources will appear softer in the 
observed frame than in the rest frame.

%
% Figure 2
%
\begin{inlinefigure}
\psfig{figure=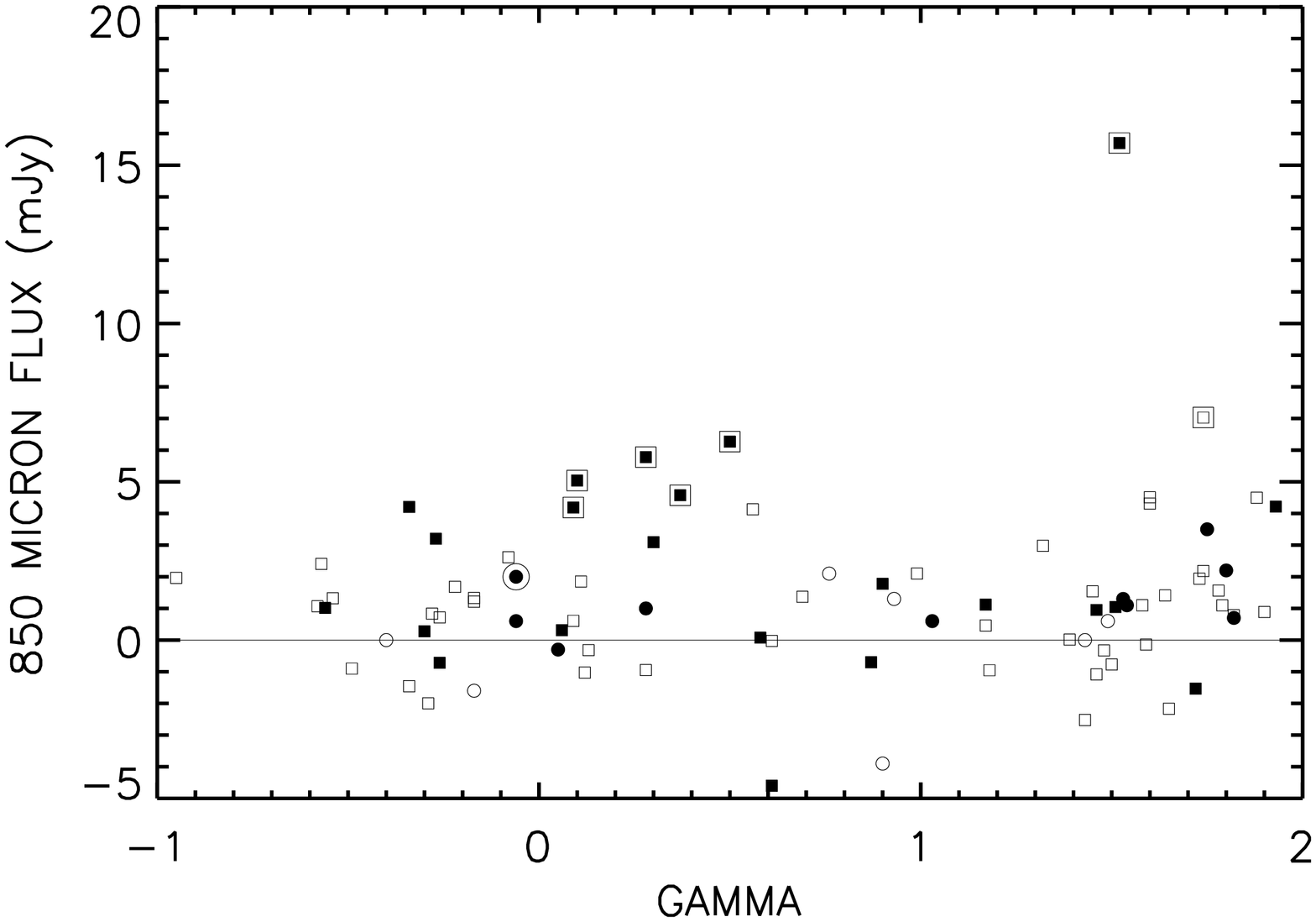,angle=90,width=3.5in}
\vspace{6pt}
\figurenum{2}
\caption{
Submillimeter flux versus $\Gamma$ for the X-ray
sources in the CDF-N (squares) and SSA13 (circles) fields
with hard X-ray fluxes above
$5.0\times 10^{-16}$~erg~cm$^{-2}$~s$^{-1}$.
Sources with $>3\sigma$ radio fluxes are denoted by filled
symbols, and sources with $>3\sigma$ submillimeter fluxes
are denoted by a second large symbol.
}
\label{fig2}
\addtolength{\baselineskip}{10pt}
\end{inlinefigure}

In Fig.~\ref{fig2} we plot 850~$\mu$m flux versus
effective photon index for the nearly complete
CDF-N hard X-ray sample (squares) above 
$5.0\times 10^{-16}$~erg~cm$^{-2}$~s$^{-1}$ ($2-8$~keV),
where the photon indices are well defined.
We include on the figure the SSA13 hard X-ray sources
(circles) from \markcite{barger01a}Barger et al.\ (2001a)
to better sample the bright X-ray flux end.
Of the eight significantly detected submillimeter sources 
which are present in this more restricted sample, six
(including the SSA13 source) are 
clustered at low $\Gamma$ values, which 
suggests they are obscured systems, while the remaining two
(including a non-radio CDF-N source)
have quite high $\Gamma$ values, which
suggests they are unobscured systems or obscured systems 
at high redshift. It should be kept 
in mind that we expect there to be one or two spurious detections
in the sample, so these objects could be false positives.

A more robust result can be obtained by considering
the average properties of the sources. Dividing the 
CDF-N+SSA13 combined sample into hard sources with $\Gamma<1$
and soft sources with $\Gamma>1$, we find that the hard
sources have an average submillimeter flux of $1.77\pm 0.21$~mJy,
which is substantially larger than the value of
$0.89\pm 0.24$~mJy in the soft sample. This suggests
that most of the submillimeter flux is indeed              
coming from the obscured sources, although even the soft
sources have a significant submillimeter excess.

In Fig.~\ref{fig3} we plot redshift versus submillimeter
flux for the unrestricted CDF-N (squares) and SSA13 (circles) 
samples. Sources with spectroscopic redshifts are separated 
from sources without by two
dotted lines at $z=0$ and at $z=3.5$. Below the $z=0$ line,
we plot the spectroscopically unidentified sources with 
magnitudes $R<24.1$, which corresponds to the $I<23.5$ definition
of Barger et al.\ (2001a) for optically bright galaxies.
These sources, although bright enough for 
spectroscopic identification, have not yet been observed; 
they will likely populate the same region of the plot 
as the low redshift sources. Above the $z=3.5$ line, we plot 
the spectroscopically unidentified sources with magnitudes
$R\ge 24.1$. These sources are generally too optically faint for
spectroscopic identification. All but two of the 
significantly detected submillimeter sources 
fall into this region of the plot, which is consistent
with the conclusion of previous SCUBA surveys 
that most submillimeter sources are optically faint. 
The two significantly detected submillimeter sources which 
are not optically faint have been observed spectroscopically:
the spectrum for the $z=1.013$ source,
which is also a radio source, shows strong Balmer absorption
features and [Ne\,{\sc III}] in emission, so
it appears to be a reliable identification, whereas the spectrum 
for the $z=0.555$ source, which is not also a radio 
source, looks relatively normal, and thus its identification 
with the submillimeter source could be spurious.

Again it is instructive to consider the average 
submillimeter properties of the X-ray sources, this
time according to redshift bin. 
The average submillimeter flux is significant in both
the $z=0$ to 1 ($0.80\pm 0.21$~mJy) and the $z=1$ to 3.5 
($1.09\pm 0.35$~mJy) spectroscopic redshift bins. Moreover,
the average submillimeter flux of the unidentified optically 
faint X-ray sources is highly significant ($1.68\pm 0.19$~mJy).
The average submillimeter flux of the 
unidentified optically bright X-ray sources is not significant 
($0.72\pm 0.56$~mJy) but is consistent with these sources
populating the same region of the plot as the sources with
spectroscopic redshifts.

Figure~\ref{fig3} shows how the unidentified optically faint
X-ray sources tend to separate according to 850~$\mu$m 
flux into sources with radio counterparts and sources without.
In fact, all but one of the significant SCUBA sources without a
spectroscopic identification have radio counterparts. This
result is again consistent with previous analyses of SCUBA data
which found that targeted SCUBA surveys of optically faint 
radio sources are a far more efficient means of identifying
SCUBA sources than random blank field surveys
(\markcite{bcr00}Barger et al.\ 2000;
\markcite{chapman01}Chapman et al.\ 2001). 

Quantitatively, the average submillimeter flux of the optically
faint X-ray sources with radio counterparts is
$2.86\pm 0.28$~mJy, which is significant at above the $10\sigma$ 
level. In contrast, the average submillimeter
flux of the optically faint X-ray sources without radio 
counterparts is not significant ($0.74\pm 0.25$~mJy).

Figure~\ref{fig3} illustrates one additional property 
we can learn about the X-ray sources with significant
submillimeter detections: their millimetric redshifts
(\markcite{cy00}Carilli \& Yun 2000; 
\markcite{bcr00}Barger et al.\ 2000).
We have overlaid on the data three curves which show the 
submillimeter flux dependence of a redshifted 20~$\mu$Jy,
40~$\mu$Jy, or 100~$\mu$Jy radio source whose spectral energy
distribution is assumed to be like that of the prototypical 
ULIG Arp~220. From these curves we can see that the unidentified 
optically faint sources that are significantly detected in the 
submillimeter are likely to fall in the redshift range between 
$z=1$ and 3.

%
% Figure 3
%
\begin{inlinefigure}
\psfig{figure=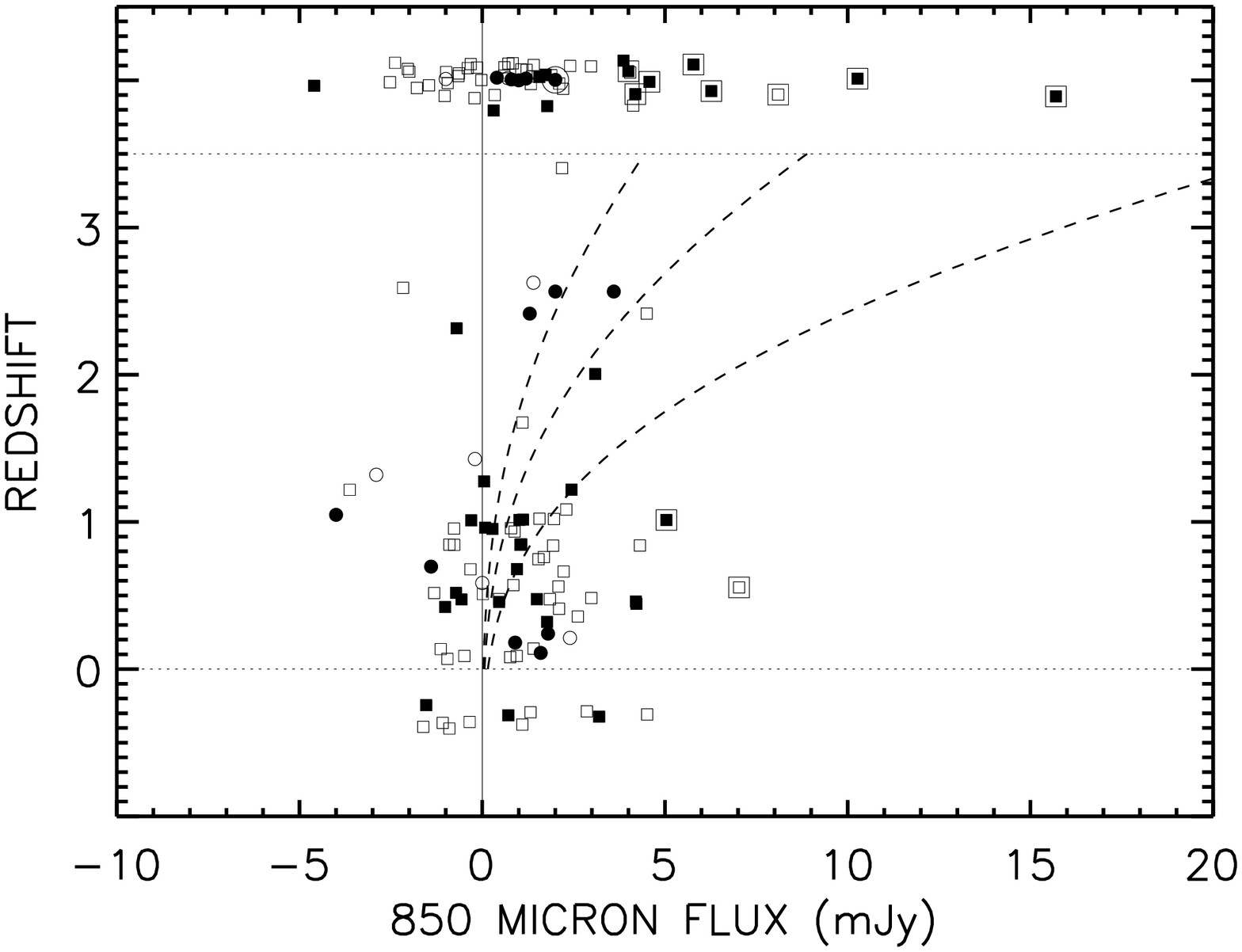,angle=90,width=3.5in}
\vspace{6pt}
\figurenum{3}
\caption{Redshift versus submillimeter flux for the X-ray
sources in the CDF-N (squares) and SSA13 (circles) fields.
Sources with $>3\sigma$ radio fluxes are denoted by filled
symbols, and sources with $>3\sigma$ submillimeter fluxes
are denoted by a second large symbol. Sources without
spectroscopic identifications are plotted at either
the bottom of the figure (the optical counterparts
have magnitudes $R<24.1$) or at the top of the figure 
(the optical counterparts have magnitudes $R\ge 24.1$),
separated from the spectroscopically identified data
by the dotted lines. Overlaid on the data are three
dashed curves which show the expected tracks of a redshifted
20~$\mu$Jy (leftmost curve), 40~$\mu$Jy, and 100~$\mu$Jy 
radio source, assuming a spectral energy distribution 
like that of the ultraluminous infrared galaxy Arp~220.
}
\label{fig3}
\addtolength{\baselineskip}{10pt}
\end{inlinefigure}

\begin{deluxetable}{ccc}
\renewcommand\baselinestretch{1.0}
\tablewidth{0pt}
\parskip=0.2cm
\tablenum{1}
\tablecaption{Average Fluxes for the CDF-N+SSA13 Sample 
in Four X-ray Flux Bins}
\small
\tablehead{
\colhead{$f(2-8$~keV) Bin} 
& Average $f(2-8$~keV) 
& Average $f(850~\mu$m) \cr
($10^{-16}$~ergs~cm$^{-2}$~s$^{-1}$) 
& ($10^{-16}$~ergs~cm$^{-2}$~s$^{-1}$) 
& (mJy) \cr
}
\startdata
$5-10$ & 7.5 & $1.7\pm 0.32$  \cr
$10-30$ & 20 & $1.8\pm 0.29$ \cr
$30-100$ & 44 & $1.3\pm 0.27$ \cr
$100-500$ & 180 & $0.25\pm 0.42$ \cr
\enddata
\label{tab1}
\end{deluxetable}

\acknowledgements
We gratefully acknowledge support from NASA through
Hubble Fellowship grant HF-01117.01-A (AJB) awarded by the
Space Telescope Science Institute, which is operated by the
Association of Universities for Research in Astronomy, Inc.,
for NASA under contract NAS 5-26555, the
University of Wisconsin Research Committee with funds
granted by the Wisconsin Alumni Research Foundation (AJB),
NSF grants AST-0084847 (AJB, PI) and AST-0084816 (LLC),
NSF CAREER award AST-9983783 (WNB), 
NASA grant NAS 8-38252 (GPG, PI),
NASA GSRP grant NGT 5-50247 (AEH),
and the Pennsylvania Space Grant Consortium (AEH).

\end{document}